\documentclass[a4paper, 12pt]{revtex4}
\usepackage[dvips]{graphicx}
\usepackage[]{indentfirst}
\usepackage[]{amsfonts}
\usepackage[]{amssymb}
\usepackage[]{bm}

\begin{document}

	\title{Autonomous learning by simple dynamical systems with a discrete-time formulation}

	\author{Agust\'in M. Bilen}
	\affiliation{Faculty of Exact and Natural Sciences, National University of Cuyo, Padre Contreras 1300, 5500 Mendoza, Argentina.} 
	
	\author{Pablo Kaluza}
	\affiliation{National Scientific and Technical Research Council \& Faculty of Exact and Natural Sciences, National University of Cuyo, Padre Contreras 1300, 5500 Mendoza, Argentina.} 
	
	\date{\today}
	
	\begin{abstract}
	We present a discrete-time formulation for the autonomous learning conjecture. The main feature of this formulation is the possibility to apply the autonomous learning scheme to systems in which the errors with respect to target functions are not well-defined for all times. This restriction for the evaluation of functionality is a typical feature in systems that need a finite time interval to process a unit piece of information. We illustrate its application on an artificial neural network with feed-forward architecture for classification and a phase oscillator system with synchronization properties. The main characteristics of the discrete-time formulation are shown by constructing these systems with predefined functions. 
	\end{abstract}

	\maketitle

	\section{Introduction}	
	The autonomous learning conjecture for the design of dynamical systems with predefined functionalities has been previously proposed by the authors \cite{kaluza_pre2014}. It extends the dynamics of a given system to a new one where the parameters are transformed to dynamical variables. The extended dynamical system then operates decreasing some cost function or error by varying the parameters through dynamics that include a delayed feedback and noise. The central feature of this idea is that the original variables and the parameters evolve simultaneously on different time scales. Because of an intimate connection with the previous publication, we do not provide here a detailed introduction and literature review, all of which can be found in Ref. \cite{kaluza_pre2014}.  
	
	The original formulation of the autonomous learning scheme works properly for systems where the cost functions are defined at all times during their evolution. However, many dynamical systems cannot satisfy this restriction since they need a finite time interval in order to process a piece of information and produce its response. As a result, the errors of such systems with respect to target functions are not defined during these processing intervals. Examples of these systems are feed-forward neural networks \cite{book_rojas}, spiking neural networks \cite{kaluza_EPJB_2014}, gene regulatory systems with adaptive responses \cite{inoue_kaneko,kaluza_inoue}, signal transduction networks \cite{kaluza_EPJB_2012}, etc. 
	
	In order to treat these kinds of systems we propose to define an iterative map for the evolution of their parameters. Between successive iterations, a dynamical system is allowed to evolve during a long enough transient time for its error function to be evaluated. With the error thus computed we update the parameters in the next iteration according to our proposed autonomous learning scheme. This approach can be seen as an adiabatic realization of the original learning scheme for continuous-time evolution.
	
	To illustrate the application of this new formulation we consider two systems: a feed-forward neural network and a Kuramoto system of phase oscillators. With the former, we have a classic example of a system that needs a time interval to process some input signal and produce its corresponding output. Since the system may classify several input patterns, the network does not only need a time to process each signal, it also needs to restart the process for each one of them. We show that this system can not only learn to classify a set of patterns but also to be robust against structural damages, namely, deletion of one of its nodes in the processing layer. In the case of the Kuramoto system, we repeat the problem of synchronization treated in our previous work \cite{kaluza_pre2014}, now using the new formalism. We then proceed to compare some aspects of the discrete and continuous-time approaches.
	
	The work is organized as follows. In section two we write down the original continuous-time autonomous learning scheme, followed by the proposed discrete-time approach. In section three we describe the two systems where we apply the new scheme. In section four we present our numerical investigations of these systems. Finally, in the last section we discuss our results and present our conclusions.

	\section{Autonomous learning theory}
	
	The continuous-time version of the autonomous learning scheme considers a simple dynamical system with $N$ variables $\bm x = (x_1, ..., x_N)$, and $Q$ parameters $\bm w = (w_1,...,w_Q$) with dynamics given by
	
	\begin{equation}
		\frac{d \boldsymbol{x}}{dt} = \boldsymbol{f}(\boldsymbol{x},\boldsymbol{w}),
		\label{equ_dynamical_system}
	\end{equation}
	
	\noindent
	and an output function $\bm{F}(\bm{x})$ which defines the task the system is responsible of executing. We further define a cost function or error $\epsilon$ between the output function and some target performance $\bm{R_0}$ as
	
	\begin{equation}
		\epsilon = |\bm{F}(\bm{x}) - \bm{R_0}|.
		\label{equ_error}
	\end{equation}
	
	Finally, to allow for a self-directed (or autonomous) minimization of the deviation $\epsilon$, we extend the original system (\ref{equ_dynamical_system}) by defining a dynamics for the parameters $\bm{w}$ as follows:
	
	\begin{equation}
		\frac{d\bm{w}}{dt} = -\frac{1}{\tau}\bm{\delta w}(t) \delta \epsilon(t) + \epsilon(t) S \bm{\xi}(t).
		\label{equ_dynamical_system_parameters} 
	\end{equation}
	
	\noindent
	In this expression $\bm{\delta w}(t) = \bm{w}(t) - \bm{w}(t - \Delta)$ and $\delta \epsilon(t) = \epsilon(t) - \epsilon(t - \Delta)$ are temporal differences at time $t$ and $t-\Delta$, where $\Delta$ is a time delay. The constant $\tau$ fixes the time scale of the evolution of $\bm{w}$. In the last term, $S$ plays the role of a noise intensity, and $\bm{\xi}(t) = (\xi_1(t), \xi_2(t),..., \xi_Q(t))$ are independent random white noises with $\langle \xi(t) \rangle = 0$ and $\langle \xi_{\alpha}(t) \xi_{\beta}(t')\rangle = 2 \delta_{\alpha \beta} \delta(t-t')$.
	
	Taken together, equations (\ref{equ_dynamical_system}) and (\ref{equ_dynamical_system_parameters}) define an extended dynamical system with combined variables $\bm x$ and $\bm w$, evolving according to two different characteristic times. In effect, if the time scale of subsystem (\ref{equ_dynamical_system}) is taken as unity and $\tau \gg 1$, the dynamics of subsystem (\ref{equ_dynamical_system_parameters}) will be slower. Additionally, the time delay $\Delta$ must satisfy $\tau \gg \Delta \gg 1$.
	
	Our conjecture is that, under an appropriate choice of $\tau$, $\Delta$ and $S$, this autonomous system will evolve along an orbit in the space of variables $\bm w$ which minimizes the system's deviation $\epsilon$ from the target performance. In \cite{kaluza_pre2014} we show that this learning method works properly for systems of oscillators where different levels of synchronization must be reached. 
	
	Equation (\ref{equ_dynamical_system_parameters}) has been designed in order to perform the weight evolution aiming to  reduce the error $\epsilon$. The interpretation of the first term on the right side of this equation is the following. If, as a consequence of the delayed feedback (memory), the system performance improves, i.e. $\delta \epsilon < 0$, with the weight correction satisfying $\delta w_i < 0$, then the weight $w_i$ should next decrease. If instead $\delta w_i > 0$, $w_i$ should increase. The opposite behavior is obtained in the case that the system performance decays with $\delta \epsilon > 0$. The second term on the right side of eq. (\ref{equ_dynamical_system_parameters}) is a noise proportional to the error $\epsilon$. Its function is to keep the dynamics from being trapped in local minima, vanishing when the error is zero.
	
	Another interpretation of eq. (\ref{equ_dynamical_system_parameters}) is that the weight evolution is controlled by a drift term (first term on the right) whose corrections are given by the memory (feedback) of the system. The second term is a stochastic exploration in parameter space proportional to the error $\epsilon$. As a result, this dynamics can be seen as a competition between a drift term of intensity $\frac{1}{\tau}$ and a stochastic term of intensity $\epsilon S$. As we will later show, a proper balance between these two terms is needed in order to obtain successful evolutions. 
	
	\subsection{Discrete-time formulation}
	
	For the application of (\ref{equ_dynamical_system}) and (\ref{equ_dynamical_system_parameters}) it is necessary that the error $\epsilon(t)$ be defined for all $t$. As we mentioned in the introduction, this restriction cannot be satisfied by systems which need a finite amount of time $T$ to execute the task given by $\bm F (\bm x)$ and, consequently, to yield a corresponding value for $\epsilon$. 
	
	Our discrete-time autonomous learning approach for the optimization of such systems considers that during each time interval $T$ the system's parameters $\bm{w}$ are fixed. After this transient, the system function $\bm F( \bm x)$ assumes a value, allowing the error function $\epsilon$ to be evaluated and the parameters $\bm{w}$ to be accordingly updated. Taking each iteration step as comprising one complete processing interval $T$, we define an iterative map analogous to eq. (\ref{equ_dynamical_system_parameters}) as:
	
	\begin{equation}
		\bm{w}(n+1) = \bm{w}(n) - K_{\tau} \bm{\delta w}(n) \delta \epsilon(n) + \epsilon(n) S \bm{\xi}(n). 
		\label{equ_discreta_parametros}
	\end{equation}
	
	\noindent
	Here, $n$ is the iteration index, $\bm{\delta w} = \bm{w}(n) - \bm{w}(n- \Delta_{\eta})$ and $\delta \epsilon = \epsilon(n) - \epsilon(n- \Delta_{\eta})$, with $\Delta_{\eta} \in \mathbb{N}$ a delay given by a certain number of iterations. The other quantities are analogous to those of eq. (\ref{equ_dynamical_system_parameters}). The constant $K_{\tau}$ determines the characteristic time for the evolution of the parameters and is equivalent to $1/\tau$. In this way, the new formulation replaces (\ref{equ_dynamical_system_parameters}) with an iterative map as the new dynamics for $\bm{w}$, and allows the system to evolve according to (\ref{equ_dynamical_system}) during an interval of time $T$ between successive iterations.

	\section{Dynamical models}
	
	In this section we present the two models we set out to design through optimization with the discrete-time autonomous learning scheme. We note that the goal of these models in this work is only to serve as examples of application and we do not pretend to analyze their properties in detail nor to compare our procedure with other methods of optimization. We focus only on the autonomous learning procedure.
	
	\subsection{Neural network model}
	
	The first example we consider is a classical feed-forward artificial neural network \cite{book_rojas} able to classify bitmaps. This kind of system is prototypic for our interest. The neural network must process different signals with a fixed set of weights, requiring a finite time interval to compute the input information and retrieve its response. During this processing interval the error is not defined, making it difficult to implement the original scheme of autonomous learning.
	
	Our model is defined as follows: a network $G$ has $N_{in}$ nodes in the input layer, $M$ nodes in the hidden layer and $N_{out}$ nodes in the output layer. A weight $w^{c,c-1}_{ij}$ is associated with a directed connection from node $j$ in layer $c-1$ to node $i$ in layer $c$. The dynamics $x_i^c$ of node $i$ in layer $c$ is
	
	\begin{equation}
		x_i^c = f \Bigg( \sum_{j=1}^{n} w^{c,c-1}_{ij} x_j^{c-1}  + w^c_{i\theta}\theta\Bigg),
		\label{equ_nn_dynamics}
	\end{equation}
	
	\noindent
	where the activation function $f$ is given by $f(x)=\tanh(x)$. For each node $i$ in layer $c$, we include a threshold value in the activation function by adding a weight $w^c_{i\theta}$ from a threshold node $\theta$ to the node in question. The threshold node is always activated with $\theta =1$.
	
	The neural network operates with a discrete-time dynamics. In the first iteration the input neurons read an input pattern $\zeta$ and get their values. In the second iteration the neurons in the hidden layer compute their state as a function of the states of the input neurons. Finally, in the third iteration the output nodes compute their states using the states of the neurons in the hidden layer. As a result, the output layer yields the response of the network after it processes the input pattern $\zeta$. 
	
	A network may repeat this $K$ times in order to compute the set of patterns $\bm \zeta = \zeta_1, ..., \zeta_K$. We thus define the error of the network as

	\begin{equation}
		\epsilon= \frac{1}{K N_{out}} \sum_{i=k}^{K} \sum_{i=1}^{N_{out}} (y^k_i - x^k_{i} )^2.
		\label{equ_nn_error}
	\end{equation}
	
	\noindent
	In this expression $y^k_i$ and $x^k_i$ are the ideal and actual responses (respectively) of the output node $i$ when the pattern $\zeta_k$ is processed by the network.

	\subsubsection{Functionality}
	
	 In order to construct functional networks, i.e., networks with small $\epsilon$, we use our new discrete-time formalism of autonomous learning. We set the iterative map (\ref{equ_discreta_parametros}) as the evolution law for the weights of the neural network. In order to use this equation we sort all the weights $w^{c,c-1}_{ij}$ and $w^{c}_{i\theta}$ in one linear array $\bm w$.
	
	\subsubsection{Robustness}
	
	We are also interested on retaining the functionality of a network when destructive mutations or damages alter the network structure. This property of structural robustness is a key feature of neural networks \cite{book_rojas, robustness_nn} and several biological systems \cite{robustness_biologicos}. Our aim is to employ the autonomous learning scheme as well to improve the robustness of a given system. The following ideas concerning robustness have already been applied to flow processing networks \cite{kaluza_pre_2007}. 
	
	Consider a network $G$ with error $\epsilon$. If we delete one if its $M$ nodes in the hidden layer, we get a new network $G_i$ with error $\epsilon_i$. In general, the damage introduced will worsen the functionality of the original network $G$, that is, we will have $\epsilon < \epsilon_i$. We say that a damaged network is no longer functional if $\epsilon_i > h$, where $h$ is thus defined as the maximum error below which a network is still considered functional. We can repeat this process of deletion for all nodes in the hidden layer to define the robustness $\rho(G)$ of network $G$ as the ratio between the number of damaged functional networks and the total number of possible damages $M$:
	
	\begin{equation}
		\rho(G) = \frac{1}{M} \sum_{i=1}^{M} \Theta[h - \epsilon_i(G)].
		\label{equ_nn_robustnez}
	\end{equation}
	
	\noindent
	Here, $\Theta$ is the Heaviside function with $\Theta(x) = 0$ if $x<0$ and $\Theta(x) = 1$ if $x \geq 0$.
	
	Since our goal is the construction of functional and robust networks, we must aim for $\epsilon \rightarrow 0$ and $\rho \rightarrow 1$. We propose the simultaneous optimization of these two quantities through the use of a biparametric autonomous learning scheme for the weights as follows:
	
	\begin{eqnarray}
		\bm w(n+1) &=& \bm w(n) -  \bm{\delta w}(n) \Big[K_{\epsilon}\delta \epsilon(n)  - K_{\rho}\delta \rho(n) \Big] \nonumber \\ 
				      & &   + \Big[ \epsilon(n) S_{\epsilon}  + \big(1-\rho(n)\big) S_{\rho} \Big]\bm{\xi}(n).
		\label{equ_nn_pesos_robustnez}
	\end{eqnarray}
	
	\noindent
	In this expression, we have two constants $K_{\epsilon}$ and $K_{\rho}$ related to the drift term, and two constants $S_{\epsilon}$ and $S_{\rho}$ related to the noise intensity. It is directly seen that this biparametric prescription for the evolution of $\bm{w}$ is essentially a superposition of (\ref{equ_discreta_parametros}) as applied individually to the optimization of $\epsilon$ and $\rho$. Note that the robustness must be calculated in each iteration, and therefore the $M$ nodes are removed one by one in each iteration.

	\subsection{Kuramoto system}
	
	We take as a second example of a dynamical system the Kuramoto model studied in our previous article \cite{kaluza_pre2014}, now applying the discrete-time autonomous learning prescription for the evolution of the coupling weights.
	
	The Kuramoto model \cite{Kuramoto} of coupled phase oscillators is described by equations

	\begin{equation}
		\frac{d \phi_i}{dt} = \omega_i + \frac{1}{N} \sum_{j=1}^{N} w_{ij} \sin(\phi_j - \phi_i)
		\label{equ_kuramoto}
	\end{equation}

	\noindent 
	where $\phi_i$ is the phase and $\omega_i$ is the natural frequency of oscillator $i$. The interactions are characterized by weights $w_{ij}$. They are symmetric, i.e.,  $w_{ij} = w_{ji}$, and can be positive or negative. 

	The synchronization of the system is quantified by the Kuramoto order parameter
	
	\begin{equation}
		r(t) = \frac{1}{N} \Bigg| \sum_{j=1}^{N} \exp(i\phi_j) \Bigg|.
		\label{equ_kuramoto_order_parameter}
	\end{equation}
	
	\noindent
	Due to time fluctuations, we work with the mean order parameter $R(t)$ defined as
	
	\begin{equation}
		R(t) = \frac{1}{T} \int_{t-T}^{t}r(t')dt', 
		\label{equ_kuramoto_mean_order_parameter}
	\end{equation}
	
	\noindent
	where $T$ is the time interval we consider for its calculation. $R(t)$ can vary between zero and one. In the case of full phase synchronization, $R(t) \to 1$.
	
	The aim of this example is to construct a system of oscillators able to autonomously learn to reach a target order parameter $P$. The error associated to this function is defined as
	
	\begin{equation}
		\epsilon(t) = |P - R(t)|.
		\label{equ_kuramoto_error}
	\end{equation}

	The discrete-time autonomous learning scheme dictates that the system (\ref{equ_kuramoto}) evolve for a time $T$ after each learning iteration. The iterative map for the weights is a variation of eq. (\ref{equ_discreta_parametros}) for our system of oscillators:

	\begin{eqnarray}
		w_{ij}(n+&1&) = w_{ij}(n) - K_{\tau}  \delta w_{ij}(n) \delta \epsilon(n)  \nonumber \\ 
				      &+&  \lambda\frac{w_{ij}(n)}{v(n)} \Big( W - v(n) \Big) + \epsilon(n) S \xi_{ij}(n).
		\label{equ_kuramoto_pesos}
	\end{eqnarray}
	
	\noindent
	Note that we compute the weights $w_{ij}$ (with $i,j = 1,...,N$) only for $j>i$, because of the symmetry of the interactions.
	
	As in our previous work, we add a term to control the total interaction in the system. The mean absolute weight $v(t)$ is defined as

	\begin{equation}
		v(n) = \frac{1}{N(N-1)} \sum_{i,j=1}^{N} | w_{ij}(n) |. 
		\label{equ_kuramoto_medio_absoluto}
	\end{equation}	
	The control parameter $\lambda$ determines how strong the redistribution of the weights is, $W$ being the ideal absolute value of $v$.

	As a result of this dynamics the system of oscillators evolves to a mean order parameter $R(t) = P$ by redistributing the weights and maintaining the total absolute value $v(t)=W$.

	\section{Numerical results}
	
	We present several numerical studies of the proposed systems. We show examples of evolutions where the autonomous learning scheme is able to lead the system to the target states and we analyze particular aspects of it which arise in connection with each particular model.

	\subsection{Feed-forward neural network}
	
\begin{figure}[!ht] 
		\begin{center}
			\resizebox{0.7\columnwidth}{!}{\includegraphics{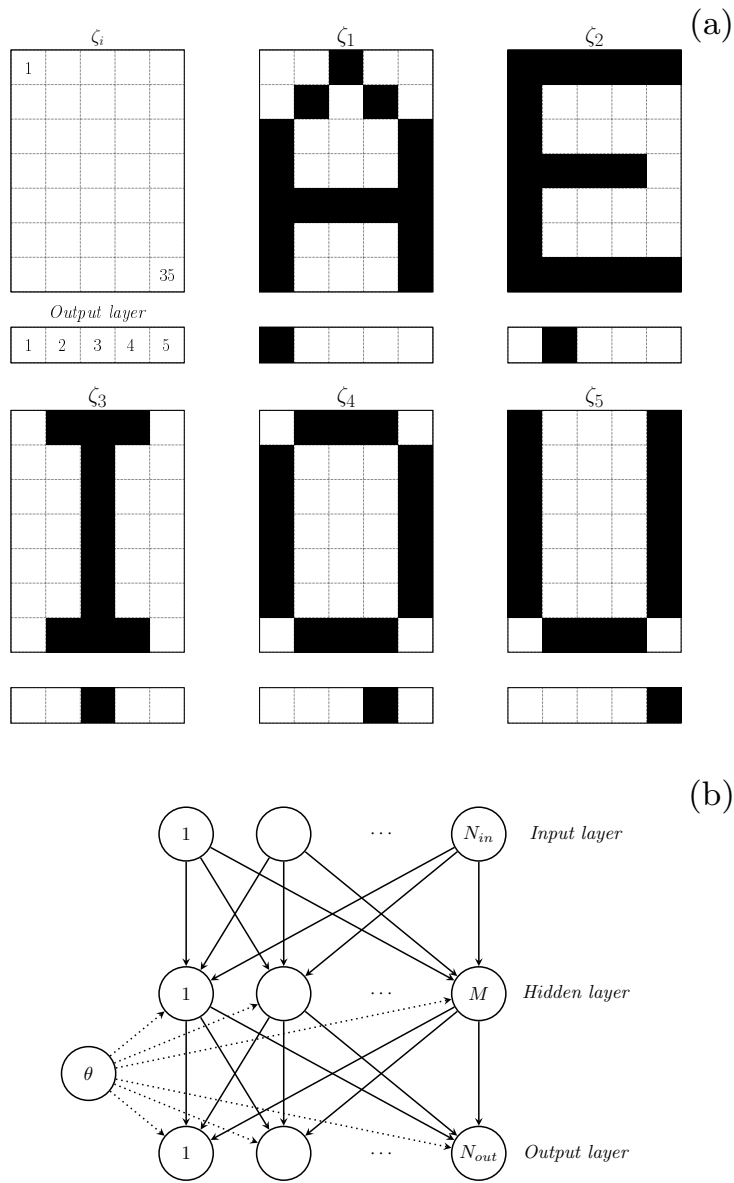}}
			\caption{(a) Set of $K=5$ input patterns $\bm \zeta$. Each one encodes a vowel as a rectangular matrix of $35$ bits. We show each pattern $\zeta_i$ and its ideal response on the output layer. A black pixel indicates activation and a white one inactivation. (b) Schematic representation of the feed-forward neural network with three layers.}
			\label{fig_figure1} 
		\end{center}
	\end{figure}

	The neural network we work with must learn to classify the vowels, i.e., there are $K=5$ input patterns $\bm \zeta$. The letters are given as matrices of binary pixels, as shown in fig. \ref{fig_figure1}.a. Each pixel acts on an input neuron, with black squares indicating activation and white ones inactivation of the associated neuron.
	
	The network has $N_{in}=35$ input neurons, one hidden layer with $M=15$ nodes and $N_{out}=5$ output neurons. A schematic representation of the network is shown in fig. \ref{fig_figure1}.b.
	
	\subsubsection{Functional networks}
	
\begin{figure}[!ht] 
		\begin{center}
			\resizebox{0.7\columnwidth}{!}{\includegraphics{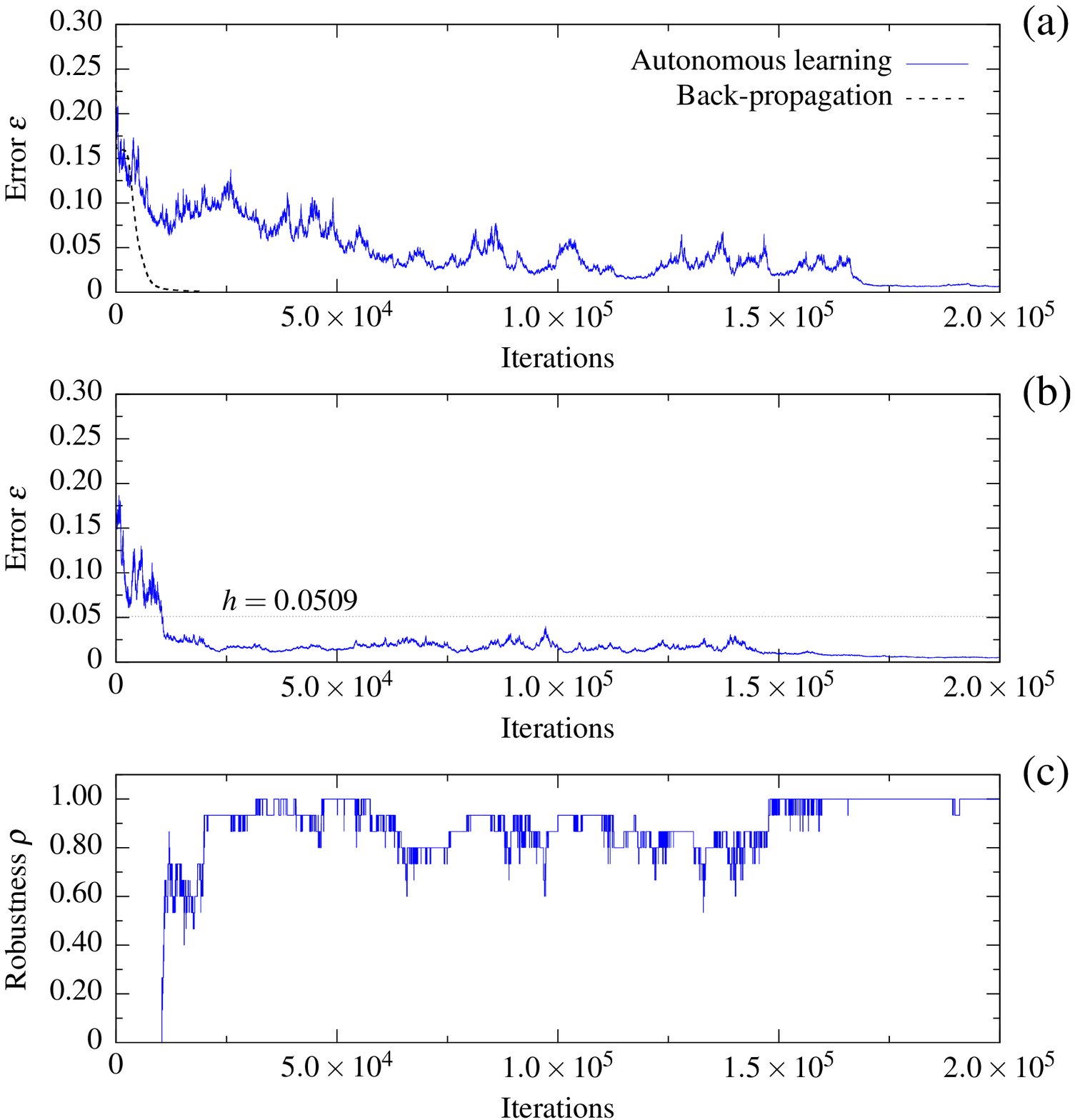}}
			\caption{ (a) Evolution of the error $\epsilon$ as a function of the number of iterations for the learning of functionality by using the discrete-time autonomous learning (blue curve) and a standard back-propagation algorithm (black dashed curve). Evolution of the error $\epsilon$ (b) and robustness $\rho$ (c) as a function of time for the biparametric learning of functionality and robustness.}
			\label{fig_figure2} 
		\end{center}
	\end{figure}

	Our first experiment consists in the realization of a full evolution through the iterative map (\ref{equ_discreta_parametros}) to construct a functional network able to classify the letters. Figure \ref{fig_figure2}.a presents a typical evolution of the error as function of the number of iterations (blue curve). We observe that the error decreases to a relatively small value at the end of the simulation, indicating a proper average classification of the patterns. The learning parameters used in the simulation were $K_{\tau}=79.43$, $S=0.13$, and $\Delta_{\eta}=1$. As the initial conditions for the weights we set $\bm w(0) = 0$ and $\bm w(-\Delta_{\eta}) = 0$. 
	
	As a matter of comparison, we add a standard back-propagation realization for a supervised learning for this system in fig. \ref{fig_figure2}a. We observe that the error as a function of the number of iterations (black dashed curve) converges to small values much faster than for the discrete-time autonomous learning scheme (blue curve). This important difference in performance is due to the deterministic character of the back-propagation algorithm in contrast with the stochastic searching of the autonomous learning. The learning factor used in the back-propagation realization was $\alpha = 0.01$.

	\subsubsection{Convergence}
	
	As we have seen from the interpretations of eqs. \ref{equ_discreta_parametros} and \ref{equ_dynamical_system_parameters}, there is a competition between the drift term controlled by $K_{\tau}$ and the stochastic term of exploration controlled by $S$. As a result, there exist certain combinations of these two values $K_\tau$ and $S$ where the learning is optimum. Generally, combinations which differ from such optimum ones result in failed evolutions where the system cannot learn.
	
	The second numerical experiment is related to finding these optimum values for $K_{\tau}$ and $S$, that is, those values for which the evolutions converge to the smallest values of $\epsilon$ in a fixed number of iterations. In order to find them, we run several simulations with ensembles of $100$ networks, fixing for each ensemble the value of $K_{\tau}$ and $S$, and evaluating the mean error $\langle \epsilon \rangle$ over the ensemble after $1 \times 10^4$ iterations. The results are shown in fig. 3a, where it can be seen that there is a clear minimum of $\langle \epsilon \rangle$ at $(\log K_{\tau} = 1.75, \log S=-0.75)$.
	
	\begin{figure}[!ht] 
		\begin{center}
			\resizebox{0.7\columnwidth}{!}{\includegraphics{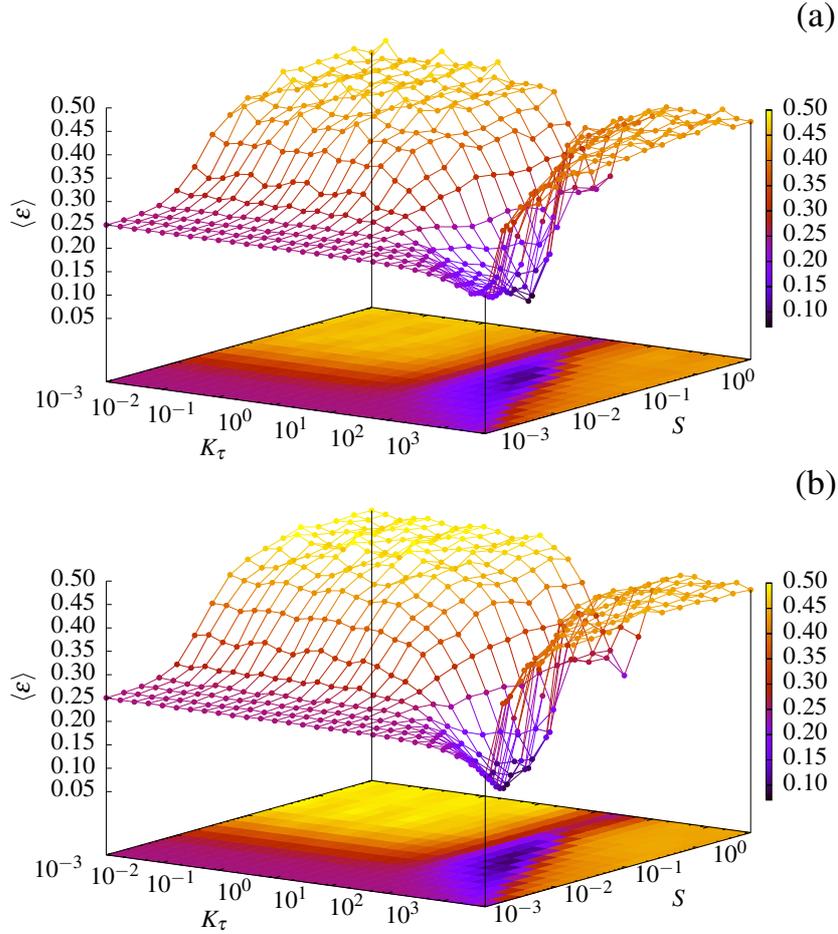}}
			\caption{Mean error $\langle \epsilon \rangle$ as a function of $K_{\tau}$ and $S$. We observe a minimum of this function at $(\log K_{\tau} = 1.75, \log S=-0.75)$, where $\langle \epsilon \rangle = 0.08$. In panel (a) we consider the classification problem only for five input pattern (only vowels), and, in panel (b) we consider this problem for $15$ input patters (vowels and digits).}
			\label{fig_figure3} 
		\end{center}
	\end{figure}
		
	The previous study considered only five input patterns for classification. Now, in order to get more robust results against the number of patterns to be classified, we consider $15$ input patterns with the same network characteristics. This set of $15$ patterns consists of the five vowels shown in fig. \ref{fig_figure1}a and the ten numerical digits from $0$ to $9$. The number of output nodes is now $15$. The mean error $\langle \epsilon \rangle$ as a function of $K_{\tau}$ and $S$ is shown in fig. \ref{fig_figure3}b. We observe that almost the same error surface is found as in the previous study, with the same optimum values for $K_{\tau}$ and $S$. 
	
	\subsubsection{Robust networks}
	
	We now consider the biparametric weight evolution given by eq. \ref{equ_nn_pesos_robustnez} that aims to minimize the error $\epsilon$ and maximize the robustness $\rho$. We set the values of $K_{\epsilon}$ and $S_{\epsilon}$ as the ones that guarantee the best convergence in the optimization of functionality, i.e., those found in the previous study. Performing a similar study to find the optimum parameters associated with the optimization of robustness, we found the minimum of $\langle 1 - \rho \rangle$ at $(\log K_{\rho}=0.75,  \log S_{\rho}=-1.75)$. For the functionality threshold we take $h=0.0509$.
	
	 An evolution for this case is shown in fig. \ref{fig_figure2}.b. We observe that at the beginning of the learning process the error $\epsilon$ is high and, as a result, the network has zero robustness. When the error is reduced and close to the threshold $h$ the learning of robustness is automatically turned on, owing to the fact that a damaged functional network has more chances to possess an error lower than the threshold as compared with a nonfunctional network with high error. As the evolution progresses, the error is kept below $h$ and the robustness increases until it reaches its optimum value. Thus, the resulting network is functional and robust.

	\begin{figure}[!ht] 
		\begin{center}
			\includegraphics[width=0.7\columnwidth, clip]{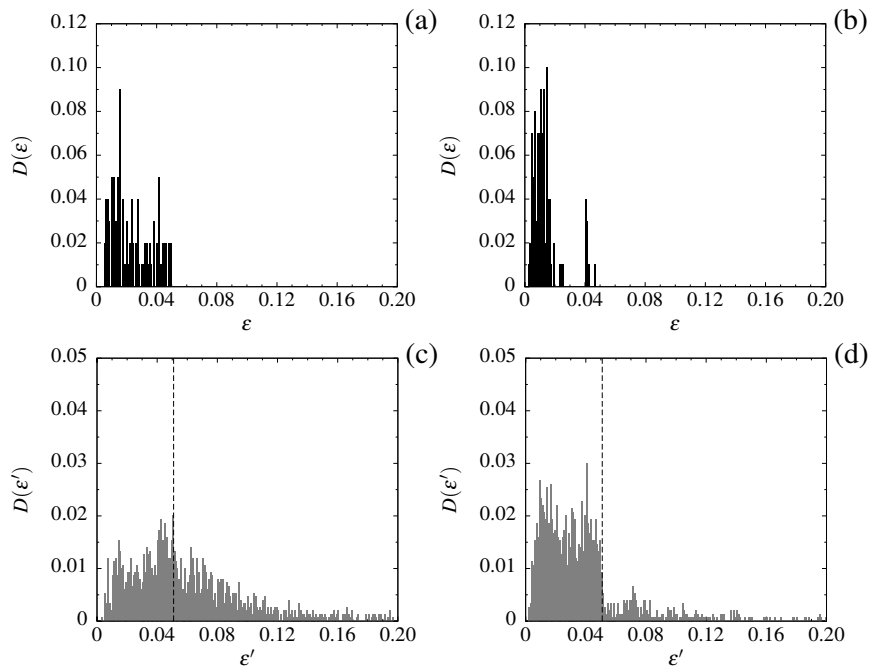}
			\caption{Histograms $D(\epsilon)$ of the errors for four ensembles of networks, all normalized to unity. (a) Histogram $D(\epsilon)$ for $100$ networks optimized only with respect to functionality. (b) Histogram $D(\epsilon)$ for $100$ networks optimized with respect to both functionality and robustness. (c) Histogram $D(\epsilon')$ for the ensemble of networks obtained by removal of hidden nodes from functional networks. (d) Histogram $D(\epsilon')$ for the ensemble of networks obtained by removal of hidden nodes from robust networks. The vertical dashed lines indicate the functionality threshold value $h=0.0509$.}
			\label{fig_figure4}  
		\end{center}
	\end{figure}
	
	In order to fix the threshold value $h$ we proceed as in article \cite{kaluza_EPJB_2012}. We optimize an ensemble of $100$ networks only by functionality (testing ensemble), computing the resulting final errors and those of the associated damaged networks. The corresponding histograms of the errors $\epsilon$ are shown in fig. \ref{fig_figure4}.a (original ensemble) and \ref{fig_figure4}.c (associated damaged ensemble) respectively. We choose $h$ as the value for which the functional ensemble has a mean robustness $\langle \rho \rangle = 0.50$. Hence, increasing the mean robustness from $\langle \rho \rangle = 0.50$ to $\langle \rho \rangle = 1$ will represent a $100\%$ increase in robustness with respect to the testing ensemble.
	
	Figures \ref{fig_figure4}.b and \ref{fig_figure4}.d show the histograms for an ensemble of $100$ networks (original and damaged, respectively) optimized to be functional \emph{and} robust through the application of the biparametric autonomous learning scheme during $2 \times 10^5$ iterations. The mean robustness of this ensemble is $\langle \rho \rangle = 0.83$. This notable increase in robustness can be clearly seen by comparing the histograms corresponding to the damaged set of networks in each case (Fig. \ref{fig_figure4}.c and \ref{fig_figure4}.d). In the case of the ensemble of networks optimized solely with respect to functionality, the error window below the threshold $h$ is much less populated than in the case for the ensemble optimized to be both functional and robust. The relative increment in robustness of the latter with respect to the testing ensemble is approximately $66\%$.

	\subsection{Kuramoto system}
	
	This system presents an interesting characteristic concerning the error function. This function is evaluated by using the mean order parameter $R(t)$ from eq. (\ref{equ_kuramoto_mean_order_parameter}), that is, a mean value over a time interval $T$. In the previous formulation for continuous-time parameter dynamics the error can be evaluated at any time. However, we are then forced to make a prescription concerning the set of weights which are most responsible for this mean value. In effect, during the interval $T$ the weights are continuously changing and it is therefore not clear which set of assumed values during $T$ are the effective ones in determining the error $\epsilon(t)$. 
	The prescription we used then was that the error $\epsilon(t)$ be related to the weights at time $t-T$, that is, we considered that the weights at the beginning of the time interval $T$ are the ones responsible for the behavior of the system at the end of the interval. Of course, we can always use a different prescription such as, for example, using the mean value of the weights over $T$. This problem has been previously considered in a different implementation of reinforcement learning for spiking neural networks \cite{Urbanczik}.
	
	A second problem related to this situation is that in the case of large enough $T$, the correlation between the weights and the error is missing. At the same time, we need a long enough period $T$ in order to minimize the variations of $R(t)$. This problem can be avoided by using our new formulation with discrete-time evolution for the parameters.
	
	We take a full connected system with $N=10$ phase oscillators with natural frequencies  $\omega_i = (i-1)/15 - 0.3$ ($i = 1,2,..., 10$). The integer delay is set to $\Delta_{\eta}=1$, and the target values to $P=0.6$ and $W=0.3$. As the initial conditions, we set $w_{ij}(0) = w_{ij}(-\Delta_{\eta}) = W$ and the initial phases $\phi_i(0)$ uniformly distributed between 0 and $2\pi$, which altogether results in the system's order parameter $R$ having an approximate initial value of $0.3$. Our aim is to increase the synchronization level to $100\%$. The system is numerically integrated using an Euler algorithm with time step $dt=0.01$. Note that between iterations of the learning algorithm the phases $\phi$ preserve their values, therefore the initial conditions are not restarted as in our previous example. We use this protocol to accelerate the simulations and to avoid transients as well as cases of multistability.
	
	\subsubsection{An evolution}
	
\begin{figure}[!ht] 
		\begin{center}
			\includegraphics[width=0.7\columnwidth, clip]{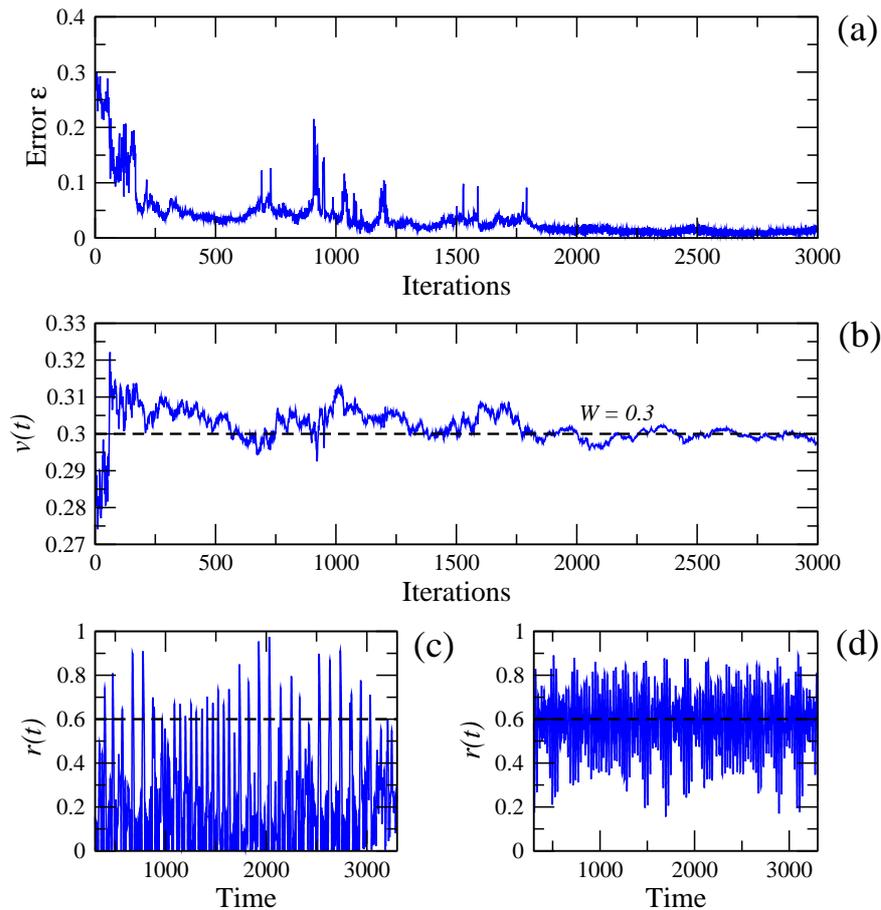}
			\caption{Example of the discrete autonomous learning for the Kuramoto model. (a) Error $\epsilon$ as a function of time. (b) Mean absolute weight $v(t)$ as a function of time. The target weight value $W=0.3$ is shown with a black dashed line. (c) and (d) Order parameters $r(t)$ as a function of time for the initial and final systems respectively. Black dashed lines show the target order parameter value $P=0.6$.}
			\label{fig_figure5} 
		\end{center}
	\end{figure}

	Figure \ref{fig_figure5} illustrates a typical successful evolution for this system. The weights of the system evolve according to eq. (\ref{equ_kuramoto_pesos}). In this simulation we use $K_{\tau}=10$, $S=0.1$ and $\lambda=0.01$. The time interval to compute the mean value $R(t)$ is $T=300$. In fig. \ref{fig_figure5}.a we plot the error as a function of the number of iterations. We observe that it decreases from $\epsilon \approx 0.3$ to $\epsilon \approx 0$ in $3000$ iterations. Not all the realizations converge to small errors in this given number of iterations. A study of the learning efficiency is presented later on in this work. 
	
	Figure \ref{fig_figure5}.b displays the mean absolute weight $v(t)$ as a function of the number of iterations. We see that at the beginning of the learning process there are relatively strong fluctuations around the target weight $W$. When the system is reaching low error values, these fluctuations vanish almost entirely as $v(t) \rightarrow W$.
		
	Figures fig. \ref{fig_figure5}.c and fig. \ref{fig_figure5}.d show the order parameter $r(t)$ as a function of time for the initial and the final systems, respectively. The simulation is performed for a time interval longer than $T$. We observe the relative improvement in the behavior of the order parameter $r(t)$ between the initial and final systems. The dashed black line shows the target value $P$ prescribed for the learning evolution. We see that, for the final system, $r(t)$ varies consistently around $P$.

	\subsubsection{Dependence on $\lambda$}
	
	Controlling the total absolute weight $v(t)$ imposes a strong restriction on the learning process. Effectively, the corresponding correcting term implies that the difference $W - v(n)$ is distributed in proportion to the strength of the connections. This way of redistributing the weights opposes their differentiation and, in general, resists the heterogeneities needed in order to find a solution. The stronger the correction by $\lambda$, the more difficult it is to reach small errors.

	\begin{figure}[!ht] 
		\begin{center}
			\includegraphics[width=0.7\columnwidth, clip]{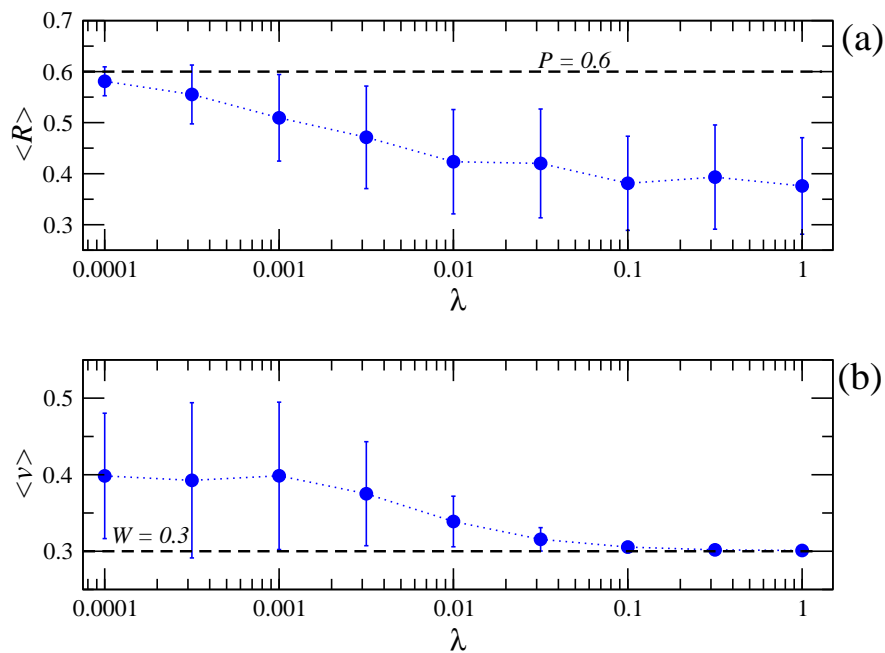}
			\caption{Mean order parameter $\langle R \rangle$ (a) and mean absolute weight $\langle v \rangle$ (b) as a function of $\lambda$ for ensembles of networks after the learning process. Errors bars indicate the dispersion of the distributions.}
			\label{fig_figure6} 
		\end{center}
	\end{figure}
	
	Figure \ref{fig_figure6}.a shows the mean order parameter $\langle R \rangle$ as a function of $\lambda$ for ensembles of $100$ networks after the learning process. Each evolution is done with $5000$ iterations, $T=200$, $K_{\tau}=10.0$ and $S=0.1$. In Figure \ref{fig_figure6}.b we show the mean absolute weight $\langle v \rangle$ of these ensembles as a function of $\lambda$. We observe that, for large values of $\lambda$, the learning scheme cannot find good solutions and the mean value $\langle R \rangle$ is far from the target value $P$. However, the mean absolute weight $\langle v \rangle$ is near its target $W$ with very small dispersion. 
	
	The opposite situation is found for small values of $\lambda$. There, the mean order parameter $\langle R \rangle$ is close to the target value $P$ with small dispersion, but the mean absolute weight $\langle v \rangle$ is much larger than $W$. As a compromise between these two tendencies, we may settle with $\lambda = 0.01$. In such a case, we find that $25\%$ of the optimized networks have  $R>0.5$. For this sub-ensemble of networks, we have $\langle R \rangle = 0.57$ and $\langle v \rangle = 0.306$. Thus, the efficiency of the learning process to find acceptable solutions is about $0.25$.

	\subsubsection{Dependence on time interval duration}
	
	We now study the dependence of the learning process on the time interval $T$. As done in the previous analysis, we optimize several ensembles of networks, this time fixing $\lambda=0.01$ and varying $T$. We keep the previous values for the rest of the learning parameters. 
	
\begin{figure}[!ht] 
		\begin{center}
			\includegraphics[width=0.7\columnwidth, clip]{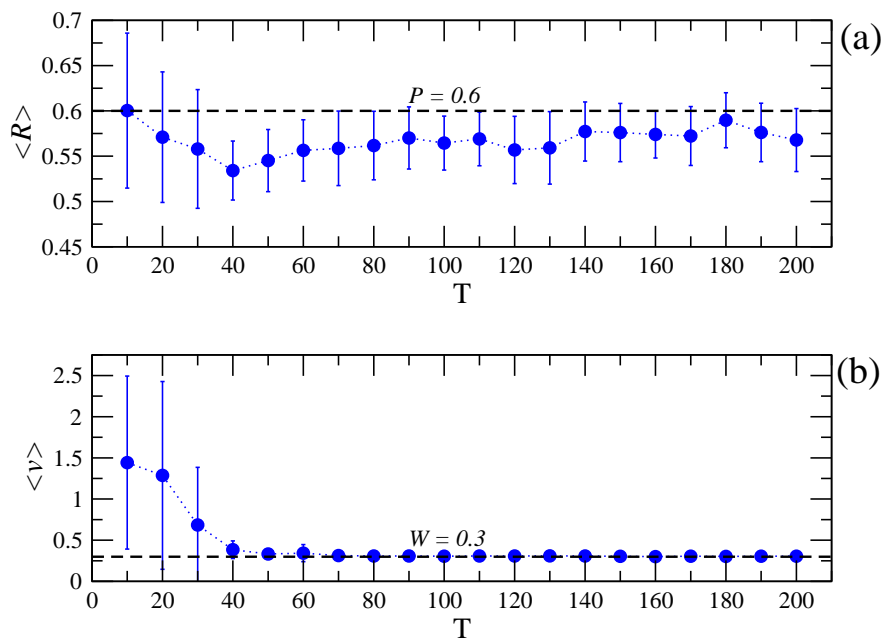}
			\caption{Mean order parameter $\langle R \rangle$ (a) and mean absolute weight $\langle v \rangle$ (b) as a function of $T$ for ensembles of networks after the learning process. Errors bars indicate the dispersion in each ensemble.}
			\label{fig_figure7} 
		\end{center}
	\end{figure}

	Figures \ref{fig_figure7}.a and \ref{fig_figure7}.b show, respectively, the mean order parameter $\langle R \rangle$ and the mean absolute weight $\langle v \rangle$ as a function $T$. The average values are computed only with successful learning cases, i.e., systems with $R>0.5$. Thus, the number of averaged systems can be different for any two points, but approximately close to $25$.
	
	We observe that, for any value of $T$, we can find networks with order parameters close to the target $P$. However, for short periods of time $T$ the fluctuations are strong and the weight restriction does not work properly. As a result, the total absolute weight $\langle v \rangle$ grows to values much larger than $W$. When we increase $T$ we observe that the weight control operates correctly and all the solutions approach $\langle v \rangle =W$. Additionally, it is interesting to note that the learning scheme works well for relatively short time windows of $T=50$, considering that in our previous work \cite{kaluza_pre2014} with the continuous-time version of autonomous learning we worked with a much longer time interval $T=200$. This result indicates that the discrete-time formulation is more efficient than the continuous version in terms of convergence speed. This fact can be understood by considering that the weights values are fixed during the interval $T$ in the new formulation and the error can be better estimated that way.

	\section{Conclusions and discussions}
	
	In this work we presented a new formulation for the autonomous learning scheme by defining an iterative map for the evolution of the parameters of a dynamical system. The utility of this discrete-time formulation resides in including within the scope of application of the autonomous learning scheme those systems which need an intrinsic time interval of finite duration to process a unit amount of information and therefore cannot measure a cost function at all times during their dynamics.
	
	The first system treated, a feed-forward neural network responsible for classifying several input patterns, is a typical example of a system subject to this restriction. We showed that our learning scheme works properly for this system, with the resulting networks able to classify the assigned patterns. Furthermore, we showed that we can implement the discrete-time scheme in biparametric form by setting a double feed-back signal in the evolution prescription for the weights, allowing us to optimize the networks with respect to the error $\epsilon$ and the robustness $\rho$ simultaneously. Our results show that the final systems can in this manner improve their robustness $66\%$ with respect to a testing ensemble. 
	
	It is important to mention that from the point of view of machine learning \cite{machine_learning}, the autonomous learning scheme can be classified as a type of reinforcement learning. These methods are characterized by their slow convergence, mainly due to the stochastic exploration of parameter space. In our case, this exploration is carried out by the multiplicative noise. As a result, our method is found to converge slower than the classical back-propagation algorithm, which constitutes a form of supervised learning by gradient descent.
	
	In the second example, a system of phase oscillators, we showed that the discrete-time formulation of autonomous learning can help to avoid the inherent fluctuations of the error function generated by the dynamics of a continuous-time dynamical system.
	
	Authors acknowledge financial support from SeCTyP-UNCuyo (project M009 2013-2015) and from CONICET (PIP 11220150100013), Argentina.

\end{document}